\documentclass[preprint,aps,showpacs]{revtex4}
\usepackage{graphicx,epsf}
\usepackage{lscape}
\usepackage{citesort}
\usepackage[psamsfonts]{amssymb}

\begin{document}

\title{{\Large{\bf Decay constants and  masses  of  light tensor  mesons ($J^P =2^+$)}}}

\author{\small R. Khosravi\footnote {e-mail: rezakhosravi @ cc.iut.ac.ir}, D.
Hatami }

\affiliation {Department of Physics, Isfahan University of
Technology, Isfahan 84156-83111, Iran }

\begin{abstract}
We calculate the masses and decay constants of the light tensor mesons
with quantum numbers $J^P =2^+$ in the framework of the QCD sum
rules in the standard model. The non-perturbative contributions up
to dimension-$5$ as important terms of the operator product
expansion are considered.
\end{abstract}


\maketitle

\section{Introduction}
In the flavor $SU(3)$ symmetry, the light $p$-wave tensor
mesons with the angular momentum $L = 1$ and total spin $S = 1$ form
an $1^3P_2$ nonet.  In other words, iso-vector mesons  $a_2(1320)$,
iso-doublet states  $K^*_2 (1430)$,  and two iso-singlet mesons
$f_2(1270)$ and $f'_2 (1525)$, are building the ground state nonet
which have been experimentally known \cite{Wang,Dombrowski}.
The quark content, $q\bar{q}$ for the iso-vector and iso-doublet tensor
resonances are obvious. The iso-scalar tensor states, $f_2(1270)$ and
$f'_2(1525)$ have a mixing wave functions where mixing angle should
be small \cite{PDG1,Li}. Therefore, $f_2(1270)$ is primarily a
$(u\bar{u}+d\bar{d})/\sqrt{2}$ state, while $f'_2(1525)$ is
dominantly $s\bar{s}$ \cite{Cheng}.

Studying the light tensor mesons properties can be useful for understanding the QCD in low energy.
In this work, we plan to consider masses and decay constants of the light
tensor mesons via the  QCD sum rules (SR). The SR have been
successfully applied to a wide variety of problems in hadron physics
(for details of this method, see \cite{shifman1,shifman2}). In this
method, calculation is started with correlation function
investigated in two phenomenological and theoretical sides.
Computing the theoretical part of the correlation function via the
operator product expansion (OPE) consists to two perturbative and
non-perturbative parts that the last part is called condensate
contributions. The condensate term of dimension-$3$ is related to
the contribution of the quark-quark condensate and dimension-$4$ and
$5$ are connected to the gluon-gluon and gluon-quark condensate,
respectively. Equating two sides of correlation function, the phenomenological and theoretical, and
applying the Borel transformation to suppress the contribution of
the higher states and continuum, the physical quantities are
estimated.

The masses and decay constants of the light tensor mesons have been
calculated in the framework of the SR in different approaches
\cite{Cheng,Shifman,Aliev,Bagan}. Also, several papers have derived
decay constant of $f_2$ from the measurement of $\Gamma(f_2\to \pi \pi)$
\cite{Terazawa,Suzuki}. As a new approach in the present work, we
calculate the masses and decay constants of the light tensor mesons by
extracting the Wilson coefficients $C^{(0)}$, and $C^{(3)}$ related
to the bare loop and quark-quark correction, respectively. Our
results for the masses and decay constants of the light tensor
mesons are in consistent agreement with the mass experimental values
and decay constant predictions of other methods.
The obtained results for the masses and decay constants can be
used in calculation of the magnetic dipole moments of the light
tensor mesons \cite{Lee}.

This paper includes three sections. In section II, the method of the
SR for calculation of the masses and decay constants of the light
tensor mesons are presented. Section III is devoted to our numerical
analysis of the masses and decay constants as well as the comparison
of them with the experimental data and predicted values by other
methods.

\section{The Method}
To compute  the  decay constants and  masses  of  the tensor  mesons
using the  two-point  QCD sum  rules, we begin with the correlation
function as
\begin{equation}\label{eqn:1}
{\Pi _{\mu \nu \alpha \beta }} = i\int {{d^4}x{e^{iq(x -
y)}}\left\langle 0 \right|T[{j_{\mu \nu }}(x) j^{\dag}_{\alpha \beta
}(y)]\left| 0 \right\rangle },
\end{equation}
the current ${j_{\mu \nu }}$, responsible for production of
the tensor meson from the QCD vacuum, is:
\begin{equation}\label{eqn:2}
{j_{\mu \nu }}(x) = \frac{i}{2}\left[ \bar q_1(x) \gamma _\mu
\stackrel\leftrightarrow D_\nu(x) q_2(x) + \bar q_1(x) \gamma _\nu
\stackrel\leftrightarrow D_\mu(x) q_2(x)\right],
\end{equation}
where $q_1$ and $q_2$ are wave functions related to two quarks composed the tensor meson. Also
\begin{eqnarray*}\label{eqn:3}
\stackrel\leftrightarrow D_\mu(x)& =&
\frac{1}{2}\left[\stackrel\rightarrow D_\mu(x) - \stackrel\leftarrow
D_\mu(x)\right],\nonumber\\
\stackrel\rightarrow D_\mu(x) &=& \stackrel\rightarrow\partial_\mu(x) - i \frac{g}{2} \lambda^a \textbf{\emph{A}}^a_\mu(x), \\
\stackrel\leftarrow D_\mu(x)&=&\stackrel\leftarrow\partial_\mu(x)+ i
\frac{g}{2} \lambda^a \textbf{\emph{A}}^a_\mu(x),
\end{eqnarray*}
${\lambda ^a}{\rm{ (}}a = 1,...,8)$ are the Gell-man matrixes and
$\textbf{\emph{A}}^a_\mu(x)$ are the external (vacuum) gluon field
that in Fock-Schwinger gauge, $x^\mu \textbf{\emph{A}}^a_\mu(x) =
0$.

As noted, the correlation function is a complex function of which the real
part comprises the computations of the phenomenology and imaginary part
comprises the computations of the theoretical part (QCD). By linking these two parts
via the dispersion relation as
\begin{equation}\label{eqn:4}
\Pi(q^2) = \frac{1}{\pi}\int\frac{\mbox{Im}\Pi(s)}{s- q^2}ds,
\end{equation}
the physical quantities such  as the decay constants and
masses of the tensor mesons are calculated.  To compute the
phenomenology part of the correlation function, a complete set
of the quantum states  of mesons is inserted in the correlation function (Eq.
(\ref{eqn:1})). After performing integral over $x$ and separating the contribution of the higher
states and continuum and opting $y = 0$, we obtain:
\begin{equation}\label{eqn:5}
{\Pi _{\mu \nu \alpha \beta }} = \frac{{\left\langle 0
\right|{j_{\mu \nu }}\left| T(q) \right\rangle \left\langle T(q)
\right|{j_{\alpha \beta }}\left| 0 \right\rangle }}{{m_T^2 -
{q^2}}} + \mbox{higher states and continuum},
\end{equation}
$m_T$ is  mass of the tensor meson $T$. The matrix elements of  Eq. (\ref{eqn:5}) can
be defined as follows
\cite{Shifman,Katz}:
\begin{equation}\label{eqn:6}
\langle 0|j_{\mu \nu}| T(q)\rangle = f_T~m_T^3~\varepsilon_{\mu \nu},
\end{equation}
where $f_T$ and ${\varepsilon _{\mu \nu }}$ are the decay constant and  polarization of the tensor meson, respectively.
Inserting Eq. (\ref{eqn:6}) into Eq. (\ref{eqn:5})  and using the
relation
\begin{eqnarray*}\label{eqn:7}
\varepsilon_{\mu\nu}\varepsilon_{\alpha\beta} = \frac{1}{2}T_{\mu
\alpha}T_{\nu \beta}+\frac{1}{2}T_{\mu\beta}T_{\nu\alpha}-
\frac{1}{3}T_{\mu\nu}T_{\alpha\beta},
\end{eqnarray*}
where $T_{\mu\nu}= -g_{\mu\nu}+\frac{q_\mu q_\nu}{m_T^2}$, and
choosing a suitable independent tensor structure, we obtain:
\begin{equation}\label{eqn:9}
\Pi_{\mu\nu\alpha\beta} = \left\{\frac{1}{2}(g_{\mu\alpha}g_{\nu\beta} +
g_{\mu\beta}g_{\nu\alpha})\frac{f_T^2~m_T^6}{m_T^2-q^2}+ \mbox{other
structures}\right\} +\mbox{ higher states}.
\end{equation}

\begin{figure}[th]
\center { \epsfxsize=15cm \epsfbox{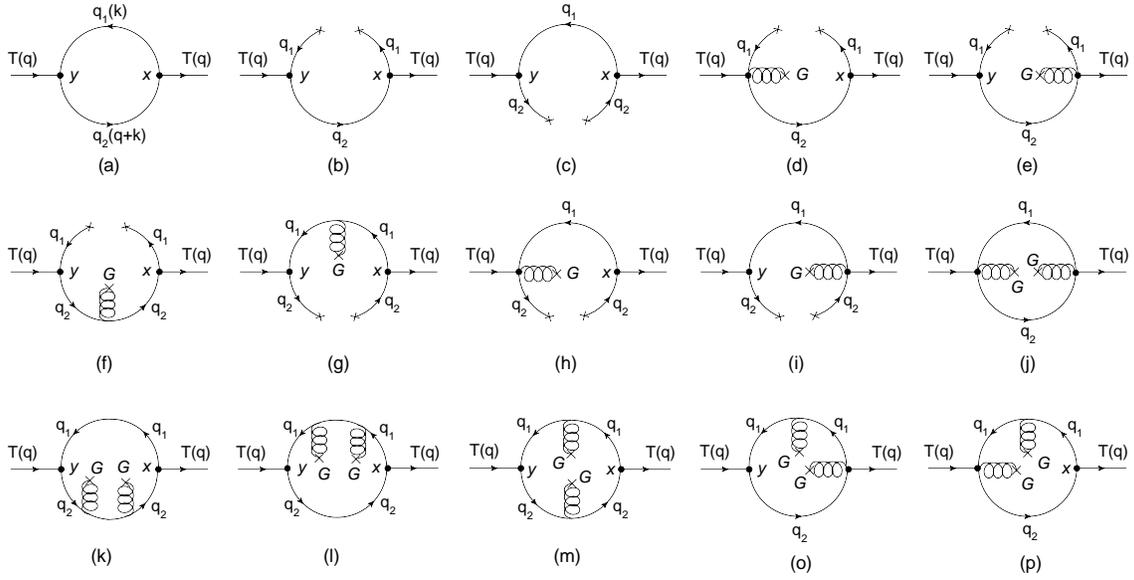} } \caption{The Fynnmans
graphs corresponding to the perturbative (a), and non-perturbative
part contributions (b,...,p), up to dimension-$5$. }\label{F1}
\end{figure}

In QCD, the correlation function can be evaluated by the operator
product expansion (OPE), in the deep Euclidean region, as
\begin{eqnarray*}\label{eqn:10}
\Pi_{\mu\nu\alpha\beta}^{QCD} = C^{(0)}_{\mu\nu\alpha\beta} +
\langle 0|\bar q q| 0\rangle C^{(3)}_{\mu\nu\alpha\beta} +\langle 0
|G_{\varphi\lambda}^a G_a^{\varphi\lambda}| 0\rangle C^{(4)}_{\mu
\nu\alpha\beta} +\langle 0| \bar q~ \sigma_{\varphi\lambda } T^a
G_a^{\varphi\lambda}~q| 0\rangle C^{(5)}_{\mu\nu\alpha\beta}+ ...,
\end{eqnarray*}
where $C^{(i)}_{\mu \nu \alpha \beta}$ are the Wilson coefficients,
$\bar q$ is the local fermion filed operator of the light quark and
$G_{\varphi \lambda }^a$ is gluon strength tensor.
The Wilson coefficients are determined by the contribution of the bare-loop, and power corrections coming
from the quark-gluon condensates of dimension-3, 4 and higher dimensions.
The diagrams corresponding to the perturbative (bare loop), and
non-perturbative part contributions up
to dimension-$5$ are depicted in Fig. \ref{F1}.

To compute the portion of the perturbative part (Fig \ref{F1}-(a)),
using the Feynman rules for the bare loop, we obtain:
\begin{eqnarray}\label{eqn:110}
C^{(0)}_{\mu\nu\alpha\beta}&=& P_{\mu\nu\alpha\beta}+P_{\mu\nu\beta
\alpha}+P_{\nu\mu\alpha\beta}+P_{\nu\mu\beta\alpha},
\end{eqnarray}
where
\begin{eqnarray}\label{eqn:11}
P_{\mu\nu\alpha\beta}&=& - \frac{i}{4}\int d^4x ~e^{iqx}~
\mbox{Tr}\left[S_{q_1}(y-x) \gamma_\mu \stackrel\leftrightarrow
D_\nu(x)
S_{q_2}(x-y) \gamma _\alpha \stackrel\leftrightarrow D_\beta(y)\right]_{|_{y=0}}.
\end{eqnarray}
Taking the partial derivative with respect to $x$ and $y$ of the
light quark free propagators, and performing the Fourier
transformation and using the Cutkosky rules, i.e.,
$\frac{1}{p^2-m^2}\rightarrow -2i\pi\delta(p^2-m^2)$,
imaginary part of the $P_{\mu\nu\alpha\beta}$ is calculated as
\begin{eqnarray}\label{eqn:12}
\mbox{Im}(P_{\mu\nu\alpha\beta})&=&- \frac{1}{{(8\pi)}^2} \int d^4k~ \delta(k^2
-m_1^2)\delta((q + k)^2 -m_2^2)~(q_\nu q_\beta + 2q_\nu
k_\beta+ 2k_\nu q_\beta +4k_\nu k_\beta)\nonumber\\
&\times& \mbox{Tr}\left[(\not k +m_1)\gamma_\mu(\not q + \not k
+m_2) \gamma _\alpha \right],
\end{eqnarray}
where $q$ is the four-momentum of the tensor meson, $m_1$ and $m_2$
are the masses  of the two quarks $q_1$, and $q_2$, respectively. To solve
the integral in Eq. (\ref{eqn:12}), we will have to deal with the
integrals as $I_{\mu\nu ...}=\int d^4k~[k_\mu k_\nu ...]
\delta(k^2-m_1^2)\delta((q + k)^2- m_2^2)$. Each integral can be
taken as an appropriate tensor structure of the $q_\mu$, $q_\nu$ and
$g_{\mu\nu}$ as
\begin{eqnarray}\label{eqn:13}
I_0&=&\frac{\pi}{2s}\sqrt{\lambda(s, m_1^2, m_2^2)} ,\nonumber \\
I_\mu&=& A (q_\mu), \nonumber \\
I_{\mu\nu}&=& B_1 (g_{\mu \nu}) +B_2  (q_\mu q_\nu), \nonumber\\
I_{\mu\nu\alpha} &=& C_1 (q_\mu q_\nu q_\alpha) +C_2 (q_\mu g_{\nu \alpha} +q_\nu g_{\mu \alpha} +q_\alpha g_{\mu \nu}),\nonumber \\
I_{\mu\nu\alpha\beta} &=& E_1 (q_\mu q_\nu q_\alpha q_\beta) +E_2
(g_{\mu \nu} g_{\alpha\beta} +g_{\mu\alpha} g_{\nu\beta} +g_{\mu
\beta} g_{\nu\alpha}) + E_3 (q_\mu q_\nu g_{\alpha \beta} + q_\mu
q_\alpha g_{\nu\beta} +q_\mu q_\beta g_{\nu\alpha} \nonumber\\&+&
q_\alpha q_\beta g_{\mu\nu} +q_\alpha q_\nu g_{\mu\beta} +q_\nu
q_\beta g_{\alpha\mu}),
\end{eqnarray}
where $s=q^2$. The quantities of the coefficient $\lambda(s, m_1^2,
m_2^2)$, $A$, $B_i,~ C_i,~ i = 1,2$, and $E_j,~j = 1,...,3$, are
stated in the appendix. By computing the trace realized in Eq.
(\ref{eqn:12}) and using the relations in Eq. (\ref{eqn:13}) and dispersion relation (to calculate the real part from the imaginary), the
perturbative part contribution of the correlation function, for the suitable structure corresponding to Eq. (\ref{eqn:9}), can
be shown as follows:
\begin{eqnarray}\label{eqn:14}
C^{(0)}_{\mu \nu \alpha \beta}&=& \frac{1}{16\pi^3}(g_{\mu \alpha
}g_{\nu \beta } + g_{\mu \beta}g_{\nu \alpha })\int_0^\infty \frac{\psi(s)}{s-q^2}ds+ \mbox{other structures},\nonumber\\
\psi(s)&=& B_1{(m_1-m_2)}^2-B_1 s-8E_2.
\end{eqnarray}

Now, the condensate terms of dimension $3, 4$ and $ 5$ are
considered. The non-perturbative part contains the quark and gluon
condensate diagrams. Our calculations show that the important
contributions coming from dimension-$3$ related to Fig \ref{F1}-(b)
and (c). The rest contributions are either zero such as (d) to (i),
or very small in comparison with the contributions of dimension-$3$
that, their contributions can be easily ignored such as (j) to (p).
It should be reminded that in the SR method, when the light quark  is a
spectator, the gluon-gluon condensate contributions are very small
\cite{Colangelo}. We will continue computation of the QCD part of
the correlation function by extracting Wilson coefficient $C^{(3)}$
corresponding to the Feynman graphs (b) and (c). For Fig.
\ref{F1}-(b), we have:
\begin{eqnarray}\label{eqn:150}
C^{(3),b}_{\mu\nu\alpha\beta}&=&N_{\mu\nu\alpha\beta}+N_{\mu\nu
\beta\alpha}+N_{\nu\mu\alpha\beta}+N_{\nu\mu\beta\alpha},
\end{eqnarray}
where
\begin{eqnarray}\label{eqn:15}
N_{\mu \nu \alpha \beta}&=&- \frac{i}{4}\int{d^4x~e^{iq x} \langle 0
| \bar q_{1\rho}(x) {\left[\gamma_\mu \stackrel\leftrightarrow
D_\nu(x) S_{q_2}(x-y)\gamma_\alpha
\stackrel\leftrightarrow D_\beta(y)\right]}_{\rho\sigma} q_{1\sigma}(y)| 0 \rangle}_{|_{y=0}}.
\end{eqnarray}
To extract the $N_{\mu \nu \alpha \beta}$,  we can expand
$q_1(x)$ around the origin as follows:
\begin{eqnarray*}\label{eqn:19}
q_1(x) &=&q_1(0) + x^\xi \stackrel\rightarrow D_\xi q_1(0) + \frac{1}{2}
x^\xi x^{\xi'} \stackrel\rightarrow D_\xi \stackrel\rightarrow
D_{\xi'} q_1(0) + \ldots.
\end{eqnarray*}
It should be noted that the Wilson coefficients are evaluated in the
deep Euclidean region ($x-y\ll 1$), and $y$ is chosen as the origin
in our calculations, therefore $x\ll 1$. Hence, we include only the
first term of this expansion in Eq. (\ref{eqn:15}). Also, using
the definition for the following matrix elements as
\begin{eqnarray*}\label{eqn:16}
\langle 0 |\bar q_{1\rho}(0)q_{1\sigma}(0)| 0 \rangle &=& \frac{1}{4}\delta_{\rho\sigma}\langle 0 |\bar q_1 q_1| 0 \rangle,\nonumber\\
\langle 0 |\bar q_{1\rho }(0)\stackrel\rightarrow D _\mu q_{1\sigma}(0)| 0 \rangle &=& - i\frac{m_1}{16}{(\gamma _\mu)}_{\rho\sigma}\langle 0 |\bar q_1 q_1| 0 \rangle,\nonumber\\
\langle 0 |\bar q_{1\rho }(0)\stackrel\leftarrow D_\mu
\stackrel\rightarrow D_\nu q_{1\sigma}(0)| 0 \rangle &=& \frac{
1}{16}(\frac{m_0^2}{2}-m_1^2 ) {(g_{\mu\nu})}_{\rho\sigma }\langle 0
|\bar q_1 q_1| 0 \rangle ,
\end{eqnarray*}
and after some calculations, we obtain:
\begin{eqnarray*}\label{eqn:17}
C^{(3),b}_{\mu\nu\alpha\beta} = \frac{1}{16}(g_{\mu\alpha}g_{\nu
\beta}+g_{\mu\beta}g_{\nu\alpha})\frac{\kappa\langle 0 |\bar q_1
q_1| 0 \rangle}{q^2- m_2^2}+ \mbox{other structures},
\end{eqnarray*}
where $\kappa=m_2(m_1^2-\frac{m_0^2}{2})$  and $m_0^2(1 GeV) = (0.8
\pm 0.2)GeV^2$ \cite{Belyaev}. After the similar calculations for Fig
\ref{F1}-(c), the final result for the non-perturbative
contributions, $C^{(3)}_{\mu \nu \alpha \beta}$ is obtained as follows:
\begin{eqnarray}\label{eqn:117}
C^{(3)}_{\mu \nu \alpha \beta} &=& \frac{1}{16}(g_{\mu\alpha }g_{\nu
\beta} +g_{\mu\beta}g_{\nu\alpha}) \left(\frac{\kappa \langle 0|\bar
q_1 q_1| 0 \rangle}{q^2 -m_2^2}+\frac{\kappa' \langle 0|\bar q_2
q_2| 0 \rangle}{q^2 -m_1^2}\right)+ \mbox{other structures},
\end{eqnarray}
where $\kappa'=m_1(m_2^2-\frac{m_0^2}{2})$.

Now, equating two the phenomenology part, Eq .(\ref{eqn:9}), and QCD part, Eqs .(\ref{eqn:14}) and (\ref{eqn:117}), of the
correlation function as well as applying the Borel transformation
\begin{eqnarray*}\label{eqn:23}
\hat B_{M^2}(q^2)\frac{1}{m^2-q^2} =\frac{1}{M^2} e^{-
\frac{m^2}{M^2}},
\end{eqnarray*}
to both sides, the decay constant of the tensor meson is computed
as
\begin{eqnarray}\label{eqn:24}
f^2_T&=& \frac{ e^{m_T^2/M^2}}{8~ m_T^6}\left\{\frac{3}{\pi^3
}\int_0^{s_T} \psi(s) e^{-s/M^2} ds \right. -\left. \kappa \langle 0
|\bar q_1 q_1| 0 \rangle e^{-m_2^2/M^2}-\kappa'
 \langle
0 |\bar q_2 q_2| 0 \rangle e^{-m_1^2/M^2}\right\},\nonumber\\
\end{eqnarray}
where $s_T$ is the continuum threshold of the tensor meson.
In the above equation, in order to subtract the contributions
of the higher states and the continuum, the quark-hadron
duality assumption is also used, i.e., it is assumed
that \cite{Colangelo}:
\begin{eqnarray}\label{eqn:230}
{\rm higher ~states}\simeq\frac{1}{\pi}\int_{s_T}^{\infty}\frac{\rho^{\rm OPE}}{s-q^2}ds,  
\end{eqnarray}
where $\rho^{\rm OPE}=\frac{3}{8\pi^2}\psi(s)$. In fact $\rho^{\rm OPE}$ is related to the Wilson coefficient $C^{(0)}_{\mu\nu\alpha\beta}$.  

Furthermore, by applying derivation to both sides of Eq.
(\ref{eqn:24}) in term of $M^2$, the mass of the tensor
meson is obtained as
\begin{equation}\label{eqn:25}
m_T^2 =\frac{\frac{3}{\pi^3}\int_0^{s_T} s~ \psi(s) e^{-s/M^2} ds
-\kappa \langle 0 |\bar q_1 q_1| 0 \rangle  ~m_2^2
e^{-m_2^2/M^2}-\kappa' \langle 0 |\bar q_2 q_2| 0 \rangle ~
m_1^2e^{-m_1^2/M^2}}{\frac{3}{\pi^3}\int_0^{s_T} \psi(s) e^{-s/M^2}
ds -\kappa \langle 0 |\bar q_1 q_1| 0 \rangle e^{-m_2^2/M^2}-\kappa'
\langle 0 |\bar q_2 q_2| 0 \rangle e^{-m_1^2/M^2}}.
\end{equation}

\section{Numerical Analysis and Conclusion}
In this section,  using Eqs. (\ref{eqn:24}) and (\ref{eqn:25}) the masses and decay constants for
the four light tensor mesons $K_2^*(1430)$, ${a_2}(1320)$,
${f_2}(1270)$, and ${f'_2}(1525)$ are computed. To
this aim, we need to insert the parameters $s_T$ and the light quark
masses in these equations. The masses of $u$ and $d$ quarks can be
numerically neglected. The mass of the $s$ quark, at the scale $1~
GeV$, is: $m_s=142~MeV$ \cite{MQH}.  The continuum threshold $s_T$ is
correlated with the energy of the first excited state of the
tensor meson under consideration. In this work, we consider the
value of the continuum threshold to be $s_T=(m_T+\Delta)^2~GeV^2$, where $\Delta=(0.20-0.30)$.
Also $\langle 0| \bar s s |0\rangle=(0.8\pm
0.2)\langle 0|\bar u u|0\rangle$, $\langle 0|\bar u u|0
\rangle=\langle 0|\bar d d|0\rangle=-(0.240\pm0.010~ GeV)^3$ that we
choose the value of the condensates at a fixed renormalization scale
of about $1~ GeV$ \cite{BLIoffe}.

\begin{figure}
\vspace{0cm}
\includegraphics[width=7cm,height=6cm]{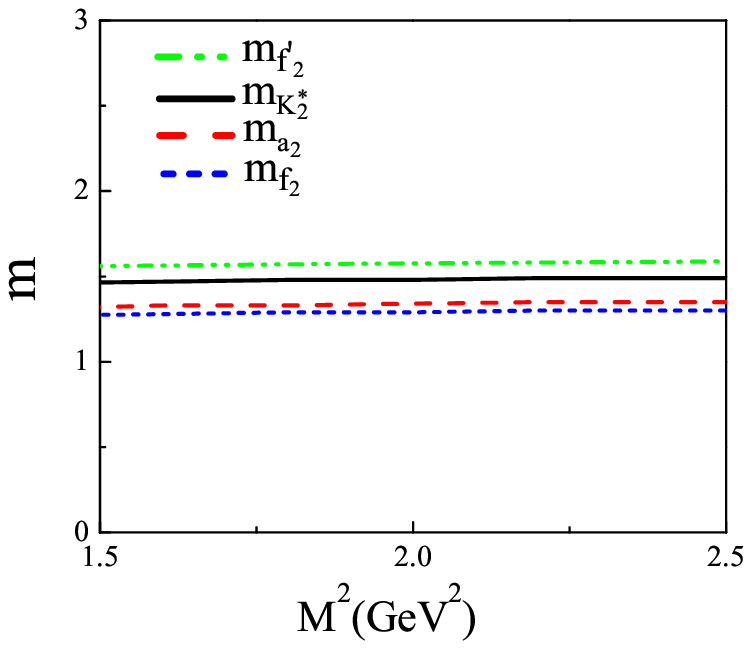}
\hspace{0.5cm}
\includegraphics[width=7cm,height=6cm]{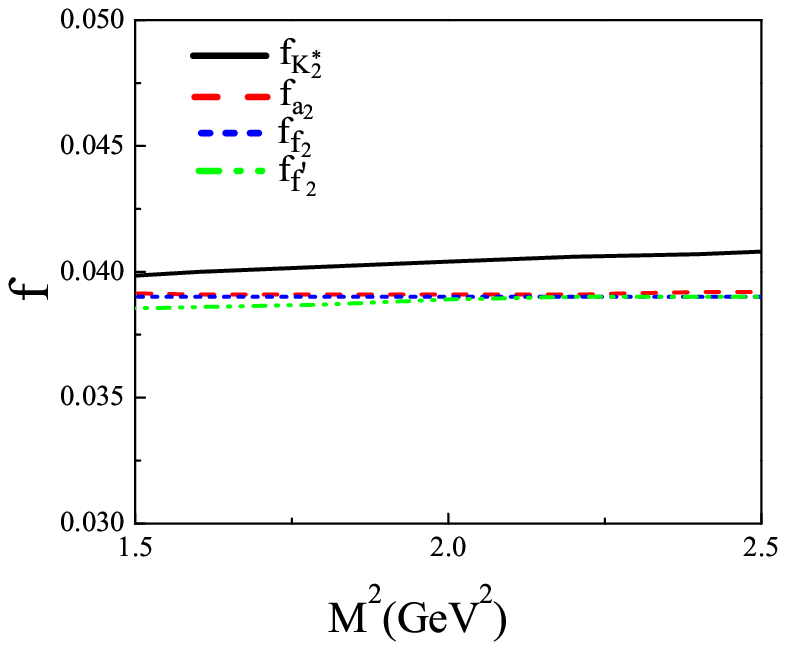}
\vspace{2ex} \caption{The dependence of the tensor
meson masses on the Borel parameter $M^2$ (left).
The same as it but for the decay constants (right).} \label{F2}
\end{figure}

The expressions for the mass and decay constant in Eqs .(\ref{eqn:23}) and (\ref{eqn:25}) contain also the Borel mass
square $M^2$ that is not physical quantities. The physical
quantities, mass and decay constant, should be independent of the
parameter $M^2$. The dependence of the masses and decay constants
of the tensor mesons on $M^2$  is shown in Fig. \ref{F2}. As can be seen
from the following graphs, in our analysis, the dependence of the
masses and the decay constants on the Broel parameter is
insignificant in the region $1.5~GeV^2 \le M^2 \le 2.5~GeV^2$.

The results of our analysis for the masses  of the
tensor mesons for different
values of $\Delta$ and $M^2=2$, are
given in Table \ref{T1}. This table contains also the experimental quantities of the light tensor mesons. As it is seen, our values for $\Delta_2=0.25 ~GeV$ are in very good
agreement with the experimental values.
\begin{table}[th]
\caption{ Comparison of the light tensor meson masses in this work
for various $\Delta$, where  $\Delta_1=0.20 ~GeV, \Delta_2=0.25 ~GeV, \Delta_3=0.30 ~GeV$, with the experimental values  in $GeV$.}\label{T1}
\begin{ruledtabular}
\begin{tabular}{cccccccccccc}
Mass&$m_T(\Delta_1)$&$m_T(\Delta_2)$&$m_T(\Delta_3)$&EXP \cite{PDG}\\
\hline
$m_{K_2^*}$&$1.39 \pm 0.22$&$1.42 \pm 0.31$&$1.46 \pm 0.42$&$1.43$\\
$m_{a_2}$&$1.28 \pm0.21$&$1.31 \pm0.30$&$1.35 \pm 0.41 $&$1.32$\\
$m_{f_2}$&$1.25 \pm0.21$&$1.28 \pm0.30$&$1.35\pm 0.41 $&$1.28$\\
$m_{f'_2}$&$1.49 \pm0.23$&$1.52 \pm0.33$&$1.55 \pm 0.45 $&$1.53$
\end{tabular}
\end{ruledtabular}
\end{table}

In Table \ref{T2}, our results for the decay constants of the tensor
mesons for different
values of $\Delta$ and $M^2=2$, as well as the obtained results via other ways in the
framework of the SR are presented. It should be noted that the decay
constant $f_T$ defined in \cite{Cheng,Bagan} differs from ours by a factor
of $1/(2m_T)$. Therefore, their values  have been rescaled and then presented in Table \ref{T2}.
\begin{table}[th]
\caption{Comparison of the decay constant values of the tensor
mesons in this work for various $\Delta$, where  $\Delta_1=0.20 ~GeV, \Delta_2=0.25 ~GeV, \Delta_3=0.30 ~GeV$, with the values obtained by others (in units of $10^{-3}$).}\label{T2}
\begin{ruledtabular}
\begin{tabular}{cccccccccccc}
Decay Constant&$f_T(\Delta_1)$&$f_T(\Delta_2)$&$f_T(\Delta_3)$&\cite{Cheng}&  \cite{Aliev}& \cite{Shifman}&\cite{Bagan}&\\
\hline
$f_{K_2^*}$&$34\pm3$&$36\pm4$&$39\pm5$&$41$&$50$&---&---\\
$f_{a_2}$& $34\pm3$& $37\pm4$&$40\pm5$&$41$&---&---&---\\
$f_{f_2}$&$35\pm3$&$38\pm4$&$41\pm6$&$40$&---&$40$&$52-72$\\
$f_{f'_2}$&$33\pm2$&$35\pm3$&$37\pm4$&$42$&---&---&$37-50$
\end{tabular}
\end{ruledtabular}
\end{table}
The results derived by us, especially for $\Delta_3=0.30 ~GeV$,  are in consistent
agreement with other values.

The errors are estimated by the variation of the Borel
parameter $M^2$, the variation of the continuum threshold
$s_T$, and uncertainties in the values of the other input parameters.
The main uncertainty comes from the continuum thresholds
of the central value, while the other uncertainties are small, constituting
a few percent.

\section*{Acknowledgments}
Partial support of Isfahan university of technology research council is appreciated.
\clearpage
\appendix
\begin{center}
{\Large \textbf{Appendix}}
\end{center}

In this appendix,  the explicit expressions of the coefficients
$\lambda(s,m_1^2,m_2^2)$, $A(s, m_1^2, m_2^2)$, $B_i(s, m_1^2,
m_2^2),~ C_i(s, m_1^2, m_2^2),~ i = 1,2$, and $E_j(s, m_1^2,
m_2^2),~j = 1,...,3$ are given.

\begin{eqnarray*}
\lambda (s,m_1^2,m_2^2) &=& {(s-m_1^2-m_2^2)}^2-4 m_1^2 m_2^2,
\quad\quad\quad\quad~ \Delta =s +m_1^2-m_2^2,
\nonumber\\
A &=& -\frac{\Delta}{2s}I_0
,\quad\quad\quad\quad\quad\quad\quad\quad\quad\quad\quad\quad~ B_1 =
\frac{I_0}{3s}\left(m_1^2 s-\frac{\Delta ^2}{4}\right),
\nonumber\\
B_2 &=& \frac{I_0}{s}\left[m_1^2 +\frac{4}{3 s}\left( \frac{\Delta
^2}{4} - m_1^2 s \right)\right], \quad\quad\quad C_1= \frac{\Delta
I_0}{s^3}\left[  \frac{1}{22}\left( \frac{\Delta ^2}{4}+ m_1^2 s
\right)-\frac{\Delta^2}{8}\right],
\nonumber\\
C_2&=& \frac{\Delta I_0}{22 s^2}\left(\frac{\Delta^2}{4}- m_1^2
s\right), \quad\quad\quad\quad\quad\quad\quad
E_1=I_0\left(\frac{\Delta^4}{12 s^4} -\frac{\Delta^2 m_1^2}{8
s^3}\right),
\nonumber\\
E_2&=&\frac{m_1^2 I_0}{36}\left(m_1^2 -\frac{\Delta ^2}{4 s}\right),
\quad\quad\quad\quad\quad\quad\quad\quad E_3=
-\frac{I_0}{72}\left(\frac{\Delta ^4}{4 s^3}+\frac{m_1^4}{s} +
\frac{5}{4}\frac{\Delta ^2 m_1^2}{s^2}\right).
\end{eqnarray*}

\end{document}